\documentclass[conference,a4paper]{APSIPA2021}
\usepackage{multirow}
\usepackage[dvips]{graphicx}
\usepackage{amsmath}
\usepackage[psamsfonts]{amssymb}
\usepackage{amsxtra}
\usepackage{threeparttable}

\usepackage{bm}
\usepackage{algorithmic,algorithm}
\usepackage{xcolor}

\usepackage{caption}
\newcommand{\vecf}[1]{\bm{#1}}
\newcommand{\matf}[1]{\bm{#1}}
\newcommand{\xeq}{\mathop{=}\limits}

\newcommand{\argmin}{\mathop{\rm arg~min}\limits}

\def\transpose{\mathsf{T}}
\def\Hermitian{\mathsf{H}}
\def\diag{\mathrm{diag}}

\def\NMF{\mathrm{NMF}}
\def\DNN{\mathrm{DNN}}
\def\ILRMA{\mathrm{ILRMA}}
\def\IDLMA{\mathrm{IDLMA}}
\def\proposed{\mathrm{prop}}
\def\R{\mathbb{R}}
\def\C{\mathbb{C}}
\def\Rp{\mathbb{R}_{\geq0}}
\def\L{\mathcal{L}}

\newcommand{\tv}[4]{\sum_{{#3}}t_{{#1}{#3},{#4}}v_{{#3}{#2},{#4}}}

\begin{document}

\title{Multichannel Audio Source Separation with Independent Deeply Learned Matrix Analysis Using Product of Source Models}

\author{
    \IEEEauthorblockN{
        Takuya Hasumi$^{\dagger}$,
        Tomohiko Nakamura$^{\dagger}$,
        Norihiro Takamune$^{\dagger}$ \\
        Hiroshi Saruwatari$^{\dagger}$,
        Daichi Kitamura$^{\star}$,
        Yu Takahashi$^{\ddagger}$,
        Kazunobu Kondo$^{\ddagger}$
    }
    \IEEEauthorblockA{
        \textit{$^{\dagger}$The University of Tokyo, Tokyo, Japan} \\
        \textit{$^{\star}$National Institute of Technology, Kagawa College, Kagawa, Japan} \\
        \textit{$^{\ddagger}$Yamaha Corporation, Shizuoka, Japan} \\
    }
}

\maketitle
\thispagestyle{empty}

\begin{abstract}
Independent deeply learned matrix analysis (IDLMA) is one of the state-of-the-art multichannel audio source separation methods using the source power estimation based on deep neural networks (DNNs).
The DNN-based power estimation works well for sounds having timbres similar to the DNN training data.
However, the sounds to which IDLMA is applied do not always have such timbres, and the timbral mismatch causes the performance degradation of IDLMA.
To tackle this problem, we focus on a blind source separation counterpart of IDLMA, independent low-rank matrix analysis.
It uses nonnegative matrix factorization (NMF) as the source model, which can capture source spectral components that only appear in the target mixture, using the low-rank structure of the source spectrogram as a clue.
We thus extend the DNN-based source model to encompass the NMF-based source model on the basis of the product-of-expert concept, which we call the product of source models (PoSM).
For the proposed PoSM-based IDLMA, we derive a computationally efficient parameter estimation algorithm based on an optimization principle called the majorization-minimization algorithm.
Experimental evaluations show the effectiveness of the proposed method.
\end{abstract}

\section{Introduction}
Multichannel audio source separation is a technique to separate concurrent sources out of mixture signals observed by a microphone array~\cite{sawada2019review}.
In the overdetermined case, many blind source separation (BSS) methods using the statistical independence between sources have thus far been proposed for decades, for example, frequency-domain independent component analysis~\cite{ikeda1999method,saruwatari2006blind} and independent vector analysis~\cite{kim2006blind,hiroe2006solution}.
One of the state-of-the-art BSS methods is independent low-rank matrix analysis (ILRMA)~\cite{kitamura2016determined}, which estimates demixing filters, using the nonnegative matrix factorization (NMF)~\cite{lee1999learning} as a source model.
When a sufficient amount of training data of the sources is available, we previously showed that the separation performance can be further improved by replacing the NMF-based source model of ILRMA with the source model based on a deep neural network (DNN)~\cite{makishima2019independent}.
We call this DNN-based method independent deeply learned matrix analysis (IDLMA), which is one of the state-of-the-art supervised but spatially blind multichannel audio source separation methods.

IDLMA uses the sourcewise DNN-based source models that are trained in advance to extract power spectrograms of target sources from noisy mixtures.
Thus, its separation performance strongly depends on the DNN-based power estimation performance.
The DNNs work well for sounds that include sources having timbres similar to those of the training data.
However, owing to the difference in musical genre and mixing, the sounds to which IDLMA is applied sometimes differ in timbre from those of the training data, which leads to performance degradation of IDLMA.
For example, in the DSD100 dataset~\cite{liutkus20172016}, which we will use in the experiments described in Section \ref{sec:experiment}, most of the sounds labeled as \texttt{bass} are played on an electric bass guitar.
The DNN trained with these sounds should work well for the electric bass guitar sounds.
However, some of the test data labeled as \texttt{bass} are played by bass instruments other than the electric bass guitar, e.g., synth bass.
This difference leads to the failure of the DNN-based source power estimation, as we will later show in Section \ref{sec:experimental-evaluation/results}.

On the other hand, since ILRMA is a fully blind method, it is free from such performance degradation caused by the timbral discrepancy.
However, the separation performance of ILRMA is often lower than that of IDLMA because its NMF-based source model assumes the low-rank structure of the source spectrograms, which does not always hold.
For the sources that do not have the low-rank structure, the DNN-based source model is effective as long as the timbral discrepancy is less significant.
From this viewpoint, the DNN- and NMF-based source models are complementary.

Motivated by this observation, we extend the DNN-based source model to encompass the NMF-based source model on the basis of the product-of-expert concept~\cite{hinton2002training}.
Since this extension combines the source models of IDLMA and ILRMA, we call it the product of source models (PoSM).
For the data to which the proposed method is applied, the DNN part represents the source components similar to the training data, whereas the NMF part represents those that appear only in the target mixture.
We build an IDLMA extension that instead uses the proposed PoSM as the source model, and we derive a computationally efficient parameter estimation algorithm using an optimization principle named the majorization-minimization (MM) algorithm~\cite{hunter2000quantile}.
Experimental results show that even for sound having a timbral gap with the DNN training data, the separation performance is further improved in the proposed method.

\section{Conventional methods}
\label{sec:conventional-methods}
\begin{figure*}[tb]
    \begin{tabular}{cc}
        \begin{minipage}{0.48\hsize}
            \centering
            \includegraphics[width=\linewidth]{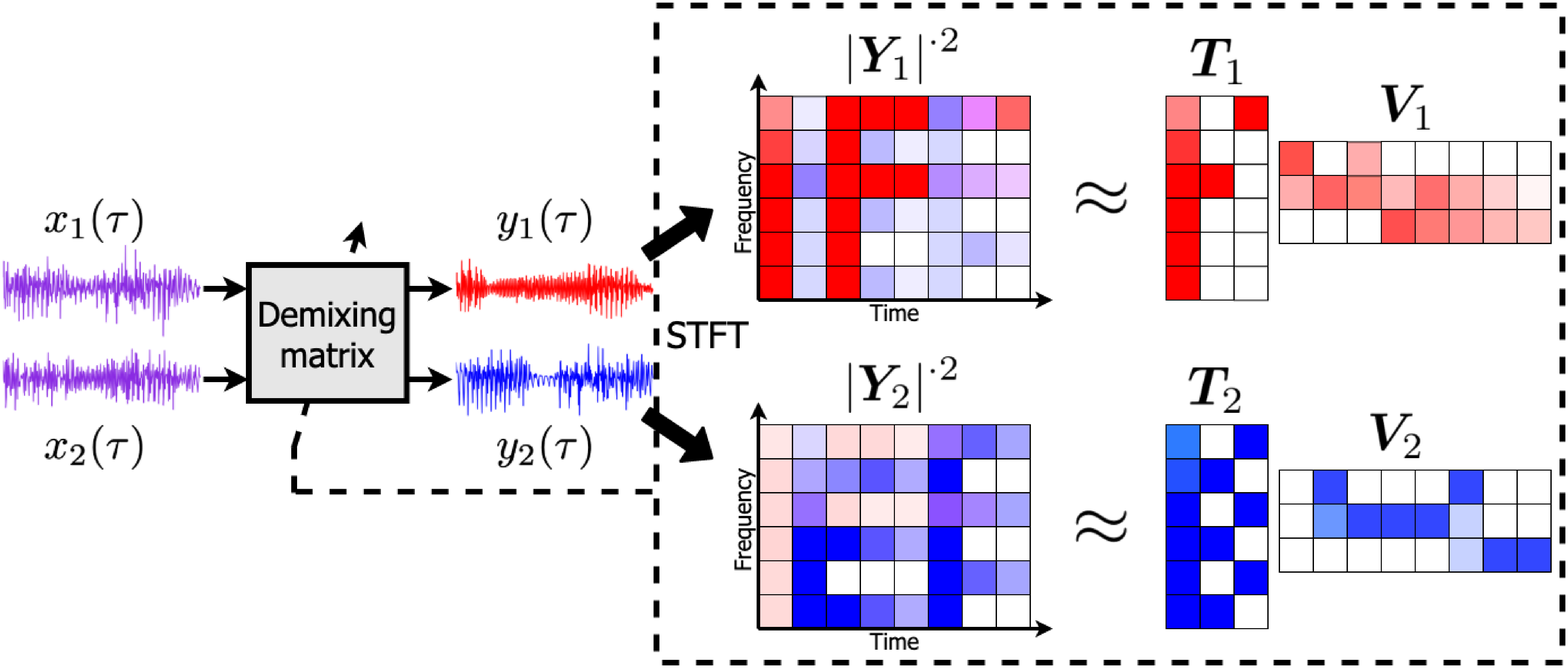}
            \text{(a) Separation process of ILRMA} 
        \end{minipage} & 
        \begin{minipage}{0.48\hsize}
            \centering
            \includegraphics[width=\linewidth]{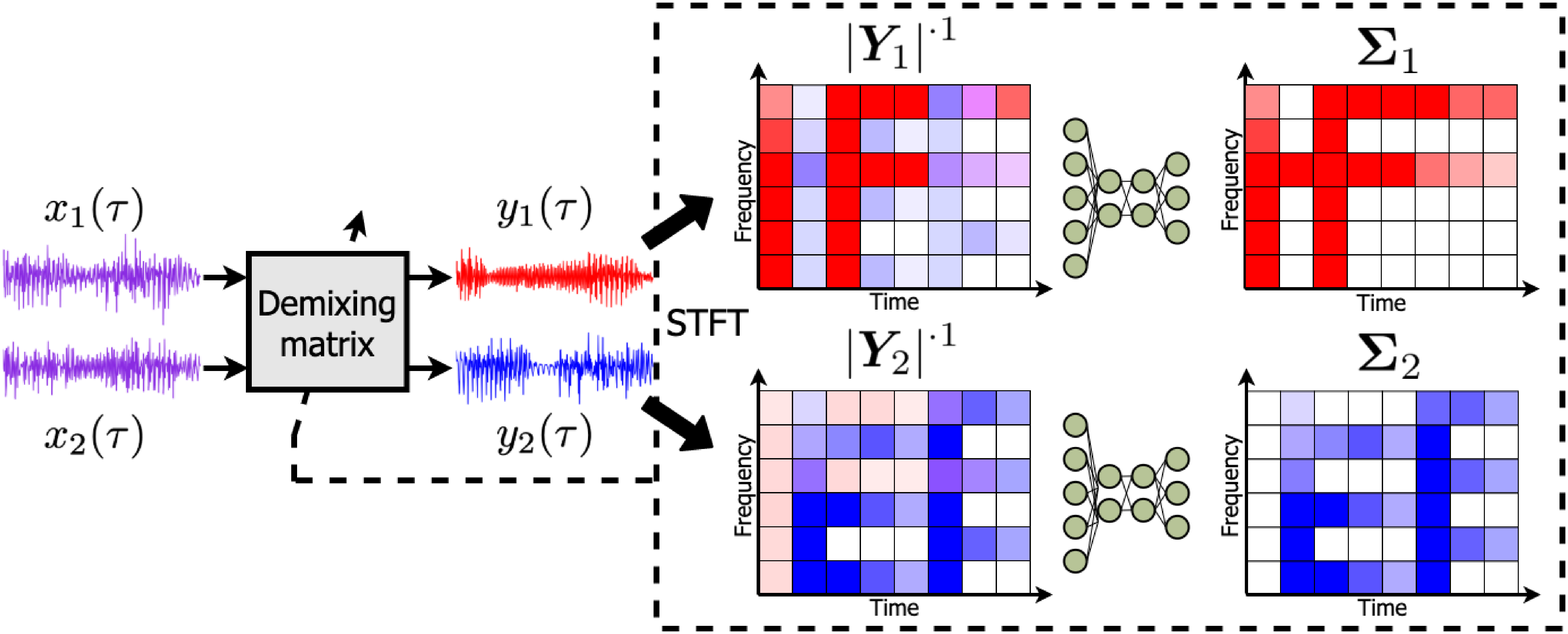}
            \text{(b) Separation process of IDLMA}
        \end{minipage}
    \end{tabular}
    \caption{Separation processes of (a) ILRMA and (b) IDLMA.}
    \label{fig:conventional-methods/overview-of_ILRMA&IDLMA}
\end{figure*}
\subsection{Formulation}
\label{sec:conventional-methods/formulation}
Let us denote the number of sources and channels by $N$ and $M$, respectively.
The short-time Fourier transform (STFT) of the source, observed, and separated signals are defined as
\begin{align}
    \vecf{s}_{ij}
    =& (s_{ij1},\ldots,s_{ijN})^{\transpose} \in\C^{N}, \\
    \vecf{x}_{ij}
    =& (x_{ij1},\ldots,x_{ijM})^{\transpose} \in\C^{M}, \\
    \vecf{y}_{ij}
    =& (y_{ij1},\ldots,y_{ijN})^{\transpose} \in\C^{N},
\end{align}
where $i=1,\ldots,I$, $j=1,\ldots,J$, $n=1,\ldots,N$, and $m=1,\ldots,M$ are the frequency, time frame, source, and channel indices, respectively, and $^{\transpose}$ denotes the transpose operator.
We define the matrices $\matf{X}_{m}\in\C^{I\times J}$ and $\matf{Y}_{n}\in\C^{I\times J}$ whose $(i,j)$th entries are $x_{ijm}$ and $y_{ijn}$, respectively.
When the mixing system is time-invariant and the window length of the STFT is sufficiently longer than the reverberation time, the observed signal $\vecf{x}_{ij}$ is represented as
\begin{align}
    \vecf{x}_{ij}
    =& \matf{A}_{i}\vecf{s}_{ij},
\end{align}
where $\matf{A}_{i}\in\C^{M\times N}$ is a mixing matrix.
If the number of channels is equal to that of sources (i.e., $M=N$) and the mixing matrix $\matf{A}_{i}$ is nonsingular, the separated signal $\vecf{y}_{ij}$ is represented as
\begin{align}
    \vecf{y}_{ij}
    =& \matf{W}_{i}\vecf{x}_{ij},
    \label{eq:conventional-methods/formulation/demixing}
\end{align}
where $\matf{W}_{i}=(\vecf{w}_{i1},\ldots,\vecf{w}_{iN})^{\mathsf{H}}\in\C^{N\times M}$ is a demixing matrix and $^{\Hermitian}$ denotes the Hermite transpose operator.

As in \cite{kitamura2016determined} and \cite{makishima2019independent}, we assume that $y_{ijn}$ follows an isotropic complex Gaussian distribution with zero mean and variance $r_{ijn}$:
\begin{align}
    p(y_{ijn};r_{ijn})
    &= \mathcal{N}_{\C}(y_{ijn};0,r_{ijn}) \notag \\
    &= \frac{1}{\pi r_{ijn}}\exp\left(-\frac{|y_{ijn}|^{2}}{r_{ijn}}\right).
    \label{eq:conventional-methods/formulation/lgm}
\end{align}
Owing to the generative model given by \eqref{eq:conventional-methods/formulation/lgm}, the problem of separating the source signals $y_{ijn}$ out of the given observed signals $x_{ijm}$ can be formulated as a maximum likelihood estimation problem with respect to $r_{ijn}$ and $\matf{W}_{i}$.
The cost function is given as the negative log-likelihood of the observed signals~\cite{sawada2019review}:
\begin{align}
    \L
    = &-\log p(\mathcal{X}) \notag \\
    = &-\log p(\mathcal{Y}) - 2J\sum_{i}\log|\det\matf{W}_{i}| \notag \\
    \xeq^{c}& \sum_{i,j,n} \left(\log r_{ijn} + \frac{|\vecf{w}_{in}^{\Hermitian}\vecf{x}_{ij}|^{2}}{r_{ijn}}\right)
    - 2J\sum_{i}\log|\det\matf{W}_{i}|,
    \label{eq:conventional-methods/cost}
\end{align}
where $\mathcal{X}=\{\matf{X}_{1},\ldots,\matf{X}_{M}\}$ and $\mathcal{Y}=\{\matf{Y}_{1},\ldots,\matf{Y}_{N}\}$ denote the sets of observed and separated signals, respectively.
Here, $\xeq^{c}$ denotes equality up to constants.
Let us denote a matrix whose $(i,j)$th entry is $r_{ijn}$ by $\matf{R}_{n}\in\Rp^{I\times J}$.
ILRMA and IDLMA are derived on the basis of the above formulation, and their difference lies in the representation of $\matf{R}_{n}$.
To distinguish the NMF- and DNN-based source models, we hereafter add superscripts $^{(\NMF)}$ and $^{(\DNN)}$ to $\matf{R}_{n}$, respectively.

\subsection{ILRMA~\cite{kitamura2016determined}}
\label{sec:conventional-methods/ILRMA}
\subsubsection{Representation of $\matf{R}_{n}^{(\NMF)}$}
ILRMA uses NMF as the source model.
The NMF represents $\matf{R}_{n}^{(\NMF)}$ as a product of two nonnegative matrices, one of which is a basis matrix $\matf{T}_{n}\in\Rp^{I\times K}$ consisting of $K$ spectral templates and the other of which is an activation matrix $\matf{V}_{n}\in\Rp^{K\times J}$ representing the temporal energies of the corresponding spectral templates.
\begin{align}
    r_{ijn}^{(\NMF)}
    =& \sum_{k}t_{ik,n}v_{kj,n}.
    \label{eq:conventional-methods/ILRMA/NMF}
\end{align}
Here, $t_{ik,n}$ and $v_{kj,n}$ are the $(i,j)$th entries of $\matf{T}_{n}$ and $\matf{V}_{n}$, respectively.
By substituting \eqref{eq:conventional-methods/ILRMA/NMF} into \eqref{eq:conventional-methods/cost}, we can write the cost function of ILRMA $\L_{\ILRMA}$ as
\begin{align}
    \L_{\ILRMA}
    \xeq^{c}& \sum_{i,j,n} \left(\log\tv{i}{j}{k}{n} + \frac{|\vecf{w}_{in}^{\Hermitian}\vecf{x}_{ij}|^{2}}{\tv{i}{j}{k}{n}}\right) \notag \\
    & - 2J\sum_{i}\log|\det\matf{W}_{i}|.
    \label{eq:conventional-methods/ILRMA/cost_ILRMA}
\end{align}

\subsubsection{Parameter Estimation Algorithm}
The parameter estimation algorithm of ILRMA iteratively updates $t_{ik,n}$, $v_{kj,n}$, and $\matf{W}_{i}$~\cite{kitamura2016determined}.
The first two terms of the cost function \eqref{eq:conventional-methods/ILRMA/cost_ILRMA} are the same form as in the cost function of NMF with the Itakura--Saito divergence criterion~\cite{fevotte2009nonnegative} up to constants.
For updating $t_{ik,n}$ and $v_{kj,n}$, we can use the convergence-guaranteed iterative algorithm derived in~\cite{fevotte2009nonnegative}, which is based on the MM algorithm~\cite{hunter2000quantile}.

The MM algorithm consists of two steps.
For a to-be-minimized cost function of $\vecf{\theta}$, $f(\vecf{\theta})$, by introducing an auxiliary variable $\bar{\vecf{\theta}}$, we construct its upper bound $f^{+}(\vecf{\theta},\bar{\vecf{\theta}})$ to satisfy the condition that there exists $\bar{\vecf{\theta}}$ such that $f^{+}(\vecf{\theta},\bar{\vecf{\theta}})$ is tangent to $f(\vecf{\theta})$ for any $\vecf{\theta}$ as follows:
\begin{align}
    \min_{\bar{\vecf{\theta}}}f^{+}(\vecf{\theta},\bar{\vecf{\theta}})
    &= f(\vecf{\theta}).
    \label{eq:conventional-methods/ILRMA/condition_tangent}
\end{align}
If $f^{+}$ can be minimized with $\vecf{\theta}$ and $\bar{\vecf{\theta}}$ in closed form, we iteratively update $\vecf{\theta}$ and $\bar{\vecf{\theta}}$:
\begin{align}
    \bar{\vecf{\theta}}
    &\leftarrow \argmin_{\bar{\vecf{\theta}}}f^{+}(\vecf{\theta},\bar{\vecf{\theta}}),
    \label{eq:conventional-methods/ILRMA/update_auxiliary} \\
    \vecf{\theta}
    &\leftarrow \argmin_{\vecf{\theta}}f^{+}(\vecf{\theta},\bar{\vecf{\theta}}).
    \label{eq:conventional-methods/ILRMA/update_target}
\end{align}
Since $f^{+}$ always satisfies \eqref{eq:conventional-methods/ILRMA/condition_tangent}, $f$ is guaranteed to be nonincreasing at each iteration.

The application of the MM algorithm to the minimization of \eqref{eq:conventional-methods/ILRMA/cost_ILRMA} with respect to $t_{ik,n}$ and $v_{kj,n}$ yields the following update rules~\cite{kitamura2016determined}:
\begin{align}
    t_{ik,n}
    & \leftarrow t_{ik,n}\left[{\frac{\displaystyle\sum_{j}\frac{v_{kj,n}}{\left(\tv{i}{j}{k'}{n}\right)^{2}}|y_{ij}|^{2}}{\displaystyle\sum_{j}\frac{v_{kj,n}}{\tv{i}{j}{k'}{n}}}}\right]^{\frac{1}{2}},
    \label{eq:conventional-methods/ILRMA/update_t} \\
    v_{kj,n}
    & \leftarrow v_{kj,n}\left[{\frac{\displaystyle\sum_{i}\frac{t_{ik,n}}{\left(\tv{i}{j}{k'}{n}\right)^{2}}|y_{ij}|^{2}}{\displaystyle\sum_{i}\frac{t_{ik,n}}{\tv{i}{j}{k'}{n}}}}\right]^{\frac{1}{2}}.
    \label{eq:conventional-methods/ILRMA/update_v}
\end{align}
By using $t_{ik,n}$ and $v_{kj,n}$ obtained from \eqref{eq:conventional-methods/ILRMA/update_t} and \eqref{eq:conventional-methods/ILRMA/update_v}, we update $r_{ijn}^{(\NMF)}$ in accordance with \eqref{eq:conventional-methods/ILRMA/NMF}.

Since \eqref{eq:conventional-methods/ILRMA/cost_ILRMA} consists only of the quadratic and log-determinant terms in $\vecf{w}_{in}$, the iterative projection (IP) algorithm~\cite{ono2011stable} can be applied, which guarantees the nonincrease in the cost function.
This method updates $\vecf{w}_{in}$ sequentially with respect to $n$:
\begin{align}
    \vecf{w}_{in}
    & \leftarrow (\matf{W}_{i}\matf{U}_{in})^{-1}\vecf{e}_{n},
    \label{eq:conventional-methods/ILRMA/update_w1} \\
    \vecf{w}_{in}
    & \leftarrow \frac{\vecf{w}_{in}}{\sqrt{\vecf{w}_{in}^{\Hermitian}\matf{U}_{in}\vecf{w}_{in}}},
    \label{eq:conventional-methods/ILRMA/update_w2} \\
    \matf{U}_{in}
    & := \frac{1}{J}\sum_{j}\frac{1}{r_{ijn}^{(\NMF)}}\vecf{x}_{ij}\vecf{x}_{ij}^{\Hermitian},
    \label{eq:conventional-methods/ILRMA/definition-of-U}
\end{align}
where $\vecf{e}_{n}\in\R^{N}$ is a unit vector whose $n$th element is one.

To compensate for the scale uncertainty between $\vecf{w}_{in}$ and $r_{ijn}^{(\NMF)}$, the projection back (PB) technique~\cite{murata2001approach} is applied to $\vecf{y}_{ij}$.
This technique determines the scale so that the sum of the separated signals matches the observation of the reference microphone whose index is denoted by $m_{\mathrm{ref}}$:
\begin{align}
    x_{ijm_{\mathrm{ref}}}
    = \sum_{n}y_{ijn}
    = \sum_{n}\vecf{w}_{in}^{\Hermitian}\vecf{x}_{ij}.
    \label{eq:conventional-methods/ILRMA/projection-back_motivation}
\end{align}
The PB technique scales $\vecf{y}_{ij}$ as
\begin{align}
    \vecf{y}_{ij}
    \leftarrow \diag(\vecf{d}_{i})\vecf{y}_{ij},
    \label{eq:conventional-methods/ILRMA/projection-back}
\end{align}
where $\diag(\vecf{d}_{i})\in\C^{N\times N}$ has the elements of $\vecf{d}_{i}\in\C^{N}$ on the main diagonal and $0$ elsewhere, and $\vecf{d}_{i}$ is computed by
\begin{align}
    \vecf{d}_{i}
    = (\matf{W}_{i}^{\transpose})^{-1}\vecf{e}_{m_{\mathrm{ref}}}.
    \label{eq:conventional-methods/formulation/solution-of-d}
\end{align}
The outline of the ILRMA separation process is shown in Fig.~\ref{fig:conventional-methods/overview-of_ILRMA&IDLMA}(a), where $|\cdot|^{\cdot q}$ returns the absolute value of each entry of a matrix to the $q$th power.

\subsection{IDLMA~\cite{makishima2019independent}}
\label{sec:conventional-methods/IDLMA}
The cost function of IDLMA is given as \eqref{eq:conventional-methods/cost} by replacing $r_{ijn}$ with $r_{ijn}^{(\DNN)}$.
In contrast to ILRMA, $\matf{R}^{(\DNN)}_{n}$ is updated using the pretrained DNNs.
The DNN for $n$th source, $\DNN_{n}$, takes the magnitude spectrogram of the current separated signal $\matf{Y}_{n}$ and outputs the estimate of the source standard deviation $\matf{\Sigma}_{n}\in\Rp^{I\times J}$:
\begin{align}
    \matf{\Sigma}_{n}=\DNN_{n}(|\matf{Y}_{n}|^{\cdot 1}).
    \label{eq:conventional-methods/IDLMA/estimation_IDLMA}
\end{align}
$\matf{R}_{n}^{(\DNN)}$ is updated as
\begin{align}
    r_{ijn}^{(\DNN)} \leftarrow \max(\sigma_{ijn}^{2},~\varepsilon),
    \label{eq:conventional-methods/IDLMA/flooring_IDLMA}
\end{align}
where $\sigma_{ijn}$ denotes the $(i,j)$th entry of $\matf{\Sigma}_{n}$, and $\varepsilon$ is a small value to avoid numerical instability.
The demixing matrix $\matf{W}_{i}$ can be updated with the IP algorithm, as in ILRMA.
To reduce the linear distortion, the PB technique is also applied to $\vecf{y}_{ij}$.
The outline of the IDLMA separation process is shown in Fig.~\ref{fig:conventional-methods/overview-of_ILRMA&IDLMA}(b).

The DNNs are trained in advance to extract the target source spectrogram $\tilde{\matf{S}}_{n}\in\C^{I\times J}$ from the single-channel instantaneous noisy mixture $\tilde{\matf{Y}}_{n}\in\C^{I\times J}$.
The loss function of training $\DNN_{n}$ is defined as
\begin{align}
    \mathcal{L}_{\IDLMA}^{(\DNN_{n})}
    &= \sum_{i,j}
    \left(\frac{|\tilde{s}_{ijn}|^{2}+\delta}{\hat{\sigma}_{ijn}^{2}+\delta}
    - \log\frac{|\tilde{s}_{ijn}|^{2}+\delta}{\hat{\sigma}_{ijn}^{2}+\delta}-1\right),
    \label{eq:conventional-methods/IDLMA/cost_DNN}
\end{align}
where $\hat{\sigma}_{ijn}\in\Rp$ and $\tilde{s}_{ijn}\in\C$ denote the $(i,j)$th entry of $\DNN_{n}(|\tilde{\matf{Y}}_{n}|^{\cdot 1})$ and $\tilde{\matf{S}}_{n}$, respectively, and $\delta$ is a small value to prevent division by zero.
Since \eqref{eq:conventional-methods/IDLMA/cost_DNN} is given by replacing $|\vecf{w}_{in}^{\Hermitian}\vecf{x}_{ij}|$ with $|\tilde{s}_{ijn}|$ in \eqref{eq:conventional-methods/cost} up to constants, the minimization of \eqref{eq:conventional-methods/IDLMA/cost_DNN} with respect to $\hat{\sigma}_{ijn}$ can be seen as a simulation of the maximum likelihood estimation with respect to $r_{ijn}^{(\DNN)}$ based on the IDLMA cost function.

\section{Proposed method}
\label{sec:proposed-method}
\subsection{Motivation}
\label{sec:proposed-method/motivation}
Since IDLMA uses trained DNNs, its separation performance is affected by the gap in timbre between the DNN training data and the sounds to which IDLMA is applied.
Such a timbral gap frequently appears in music audio signals owing to differences in mixing, musical styles, and genres.
For example, as a bass instrument, the electric bass guitar is typically used in rock and pop music, whereas the synth bass is frequently used in electronic music.
Although the sounds played by the electric and synth basses are both labeled as \texttt{bass} in the DSD100 dataset~\cite{liutkus20172016}, their spectral characteristics differ significantly, which can lead to the performance degradation of IDLMA, as we will show later in Section \ref{sec:experiment}.
Although one method to address this problem would be to collect various instrument sounds so that the trained DNN can deal with any possible timbral variations, this can be costly and impractical.

\begin{figure*}[tb]
    \centering
    \includegraphics[width=\linewidth]{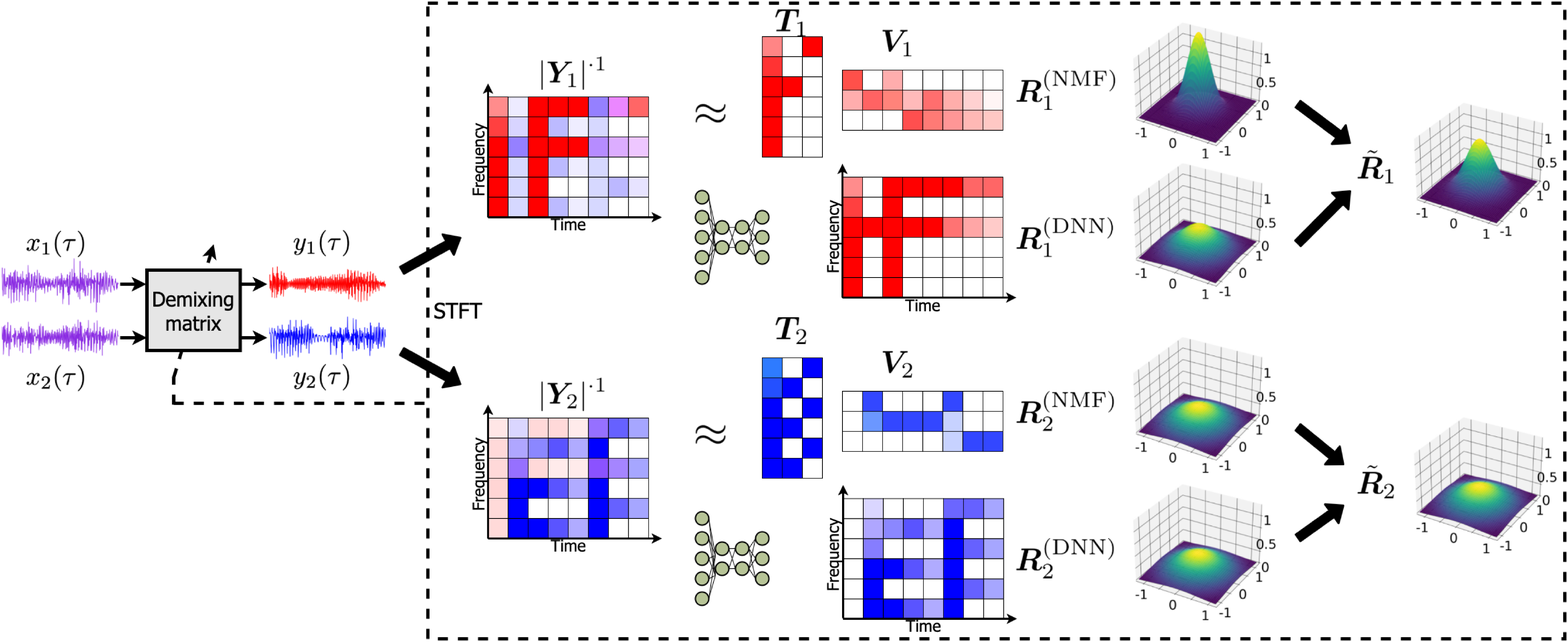}
    \caption{Separation process of proposed method.}
    \label{fig:proposed-method/NMF-updates/overview-of-proposed}
\end{figure*}
To overcome this problem, we take an approach that combines the supervised and unsupervised source models.
The former represents the source components that can be learned from the training data, and the latter represents those unique in the data to which the method is applied.
As the supervised and unsupervised source models, we can use the DNN- and NMF-based source models of IDLMA and ILRMA, respectively.
By combining them in an appropriate manner, we extend IDLMA to work robustly against the discrepancy in timbre with the training data.

\subsection{Formulation of Product of Source Models}
\label{sec:proposed-method/formulation}
Since both the DNN- and NMF-based source models are given as the generative models of the observed signals, we can combine them, following the product-of-expert concept~\cite{hinton2002training}.
This concept represents a probability distribution as a product of multiple probability distributions called experts.
Each expert corresponds to the desired constraint, and the resulting probability distribution becomes high at the events where all the constraints tend to be satisfied.

Following the product of experts, we can define a novel source model, which we call the PoSM, on the basis of $r_{ijn}^{(\NMF)}$ and $r_{ijn}^{(\DNN)}$ as
\begin{align}
    & p(y_{ijn};r_{ijn}^{(\NMF)},r_{ijn}^{(\DNN)}) \notag \\
    & \hspace{-1mm}\propto \left[\mathcal{N}_{\C}(y_{ijn};0, r_{ijn}^{(\NMF)})\right]^{\alpha}\left[\mathcal{N}_{\C}(y_{ijn};0, r_{ijn}^{(\DNN)})\right]^{\beta},
    \label{proposed-method/formulation/product-of-experts}
\end{align}
where $\alpha>0$ and $\beta>0$ are the weights for the NMF- and DNN-based probability distributions.
Since a product of two complex Gaussian distributions with zero means is also a complex Gaussian distribution with zero mean, the PoSM can be explicitly described as
\begin{align}
    p(y_{ijn};r_{ijn}^{(\NMF)},r_{ijn}^{(\DNN)})
    & = \mathcal{N}_{\C}(y_{ijn};0, \tilde{r}_{ijn}) \notag \\
    & = \frac{1}{\pi\tilde{r}_{ijn}}\exp\left(-\frac{|y_{ijn}|^{2}}{\tilde{r}_{ijn}}\right),
\end{align}
where 
\begin{align}
    \frac{1}{\tilde{r}_{ijn}}
    = \frac{\alpha}{r_{ijn}^{(\NMF)}}+\frac{\beta}{r_{ijn}^{(\DNN)}}.
    \label{eq:proposed-method/formulation/definition-of-r}
\end{align}
Interestingly, when we set $\alpha+\beta=1$, $\tilde{r}_{ijn}$ is a weighted harmonic mean of $r_{ijn}^{(\NMF)}$ and $r_{ijn}^{(\DNN)}$, which means $\alpha$ and $\beta$ balance the importance of the source estimates obtained with the DNN and NMF.

We define a matrix $\tilde{\matf{R}}_{n}\in\Rp^{I\times J}$ whose $(i,j)$th entry is $\tilde{r}_{ijn}$.
As in Section \ref{sec:conventional-methods/formulation}, the separation problem based on the PoSM can be formulated as a maximum likelihood estimation problem with respect to $t_{ik,n}$, $v_{kj,n}$, $r_{ijn}^{(\DNN)}$, and $\matf{W}_{i}$, and the cost function of the proposed method is given as
\begin{align}
    \L_{\proposed}
    \xeq^{c} & -\sum_{i,j,n}\log \left(\frac{\alpha}{\tv{i}{j}{k}{n}} + \frac{\beta}{r_{ijn}^{(\DNN)}}\right) \notag \\
    & + \sum_{i,j,n}\left(\frac{\alpha}{\tv{i}{j}{k}{n}}+\frac{\beta}{r_{ijn}^{(\DNN)}}\right)|\vecf{w}_{in}^{\Hermitian}\vecf{x}_{ij}|^{2} \notag \\
    & - 2J\sum_{i}\log|\det\matf{W}_{i}|.
    \label{eq:proposed-method/formulation/cost}
\end{align}
If $\alpha=1$ and $\beta=0$, the cost function of the proposed method \eqref{eq:proposed-method/formulation/cost} reduces to that of ILRMA given by \eqref{eq:conventional-methods/ILRMA/cost_ILRMA}.
If $\alpha=0$ and $\beta=1$, the cost function of the proposed method \eqref{eq:proposed-method/formulation/cost} reduces to that of IDLMA.

\subsection{DNN Training and Update Rules}
\label{sec:proposed-method/pretraining-and-separation-algorithm}
\subsubsection{Outline of Separation Process}
Fig.~\ref{fig:proposed-method/NMF-updates/overview-of-proposed} shows the outline of the separation process of the proposed PoSM-based IDLMA.
As in ILRMA and IDLMA, the separation is achieved by iteratively updating the parameters of the PoSM and the demixing matrix.
For the PoSM parameters, $r_{ijn}^{(\DNN)}$ is updated in the same manner as IDLMA, and $t_{ik,n}$ and $v_{kj,n}$ are updated by a convergence-guaranteed algorithm, as we will show in Section \ref{sec:proposed-method/NMF-updates}.
The variance of the PoSM $\tilde{r}_{ijn}$ is then updated using the current estimates of $t_{ik,n}$, $v_{kj,n}$, and $r_{ijn}^{(\DNN)}$.
The demixing matrix is updated by the IP algorithm followed by the PB technique.

\subsubsection{DNN Training and Update Rule of $\matf{R}_{n}^{(\DNN)}$}
\label{sec:proposed-method/pretraining-and-separation-algorithm/DNN-training-and-inference}
Since $r_{ijn}^{(\NMF)}$ represents the source components independent of the training data, we can set $\alpha=0$ and $\beta=1$ during the DNN training, which reduces the cost function \eqref{eq:proposed-method/formulation/cost} to that of IDLMA, as described in Section \ref{sec:proposed-method/formulation}.
This justifies training the DNNs in the same manner as in IDLMA, using the cost function \eqref{eq:conventional-methods/IDLMA/cost_DNN}.
In the separation process, we update $\matf{\Sigma}_{n}$ and $\matf{R}_{n}^{(\DNN)}$ in accordance with \eqref{eq:conventional-methods/IDLMA/estimation_IDLMA} and \eqref{eq:conventional-methods/IDLMA/flooring_IDLMA}, respectively.

\subsubsection{Update Rules of $\matf{R}_{n}^{(\NMF)}$}
\label{sec:proposed-method/NMF-updates}
The first and second terms of \eqref{eq:proposed-method/formulation/cost} respectively include the sums over $k$ in the logarithmic function and the denominator of the fractional function.
These terms make it difficult to analytically solve the minimization of \eqref{eq:proposed-method/formulation/cost} with respect to $t_{ik,n}$ and $v_{kj,n}$.
However, we can instead derive a computationally efficient algorithm that iteratively updates $t_{ik,n}$ and $v_{kj,n}$, using the MM algorithm~\cite{hunter2000quantile}.

Focusing on the first term of \eqref{eq:proposed-method/formulation/cost}, when $\alpha$, $\beta$, $r$, $z>0$, its second-order derivate with respect to $z$ is always negative:
\begin{align}
    \frac{\partial^{2}}{\partial z^{2}}\left[-\log\left(\frac{\alpha}{z}+\frac{\beta}{r}\right)\right]
    = - \frac{\alpha r(\alpha r+2\beta z)}{z^{2}(\alpha r+\beta z)^{2}}
    < 0.
\end{align}
Since $-\log\left(\alpha/z+\beta/r\right)$ is concave, by applying the tangent inequality to the first term of \eqref{eq:proposed-method/formulation/cost}, we obtain its upper bound:
\begin{align}
    & -\log \left(\dfrac{\alpha}{\tv{i}{j}{k}{n}}+\dfrac{\beta}{r_{ijn}^{(\DNN)}}\right) \notag \\
    &\leq \dfrac{\alpha r_{ijn}^{(\DNN)}}{\alpha r_{ijn}^{(\DNN)}+\beta c_{ijn}}\dfrac{1}{c_{ijn}}\left(\tv{i}{j}{k}{n} - c_{ijn}\right) \notag \\
    & \hspace{4.5mm} - \log\left(\frac{\alpha}{c_{ijn}} + \frac{\beta}{r_{ijn}^{(\DNN)}}\right),
    \label{eq:proposed-method/NMF-updates/tangent-equality}
\end{align}
where $c_{ijn}>0$ is an auxiliary variable.
The equality of \eqref{eq:proposed-method/NMF-updates/tangent-equality} holds if and only if
\begin{align}
    c_{ijn}
    = \tv{i}{j}{k}{n}.
    \label{eq:proposed-method/NMF-updates/tangent-equality_hold}
\end{align}
Since $1/z$ is convex for $z>0$, we can apply Jensen's inequality to the second term of \eqref{eq:proposed-method/formulation/cost}:
\begin{align}
    \frac{1}{\tv{i}{j}{k}{n}}
    \leq \sum_{k}\frac{\lambda_{ijk,n}^{2}}{t_{ik,n}v_{kj,n}},
    \label{eq:proposed-method/NMF-updates/Jensen-equality}
\end{align}
where $\lambda_{ijk,n}$ is an auxiliary variable that satisfies $\lambda_{ijk,n}\geq 0$ and $\sum_{k}\lambda_{ijk,n}=1$ for all $i,j$, and $n$.
The equality of \eqref{eq:proposed-method/NMF-updates/Jensen-equality} holds if and only if
\begin{align}
    \lambda_{ijk,n}
    = \frac{t_{ik,n}v_{kj,n}}{\tv{i}{j}{k'}{n}}.
    \label{eq:proposed-method/NMF-updates/Jensen-equality_hold}
\end{align}
In summary, the auxiliary function of $\L_{\proposed}$ is given as
\begin{align}
    \L^{+}_{\proposed}
    \xeq^{c} & \sum_{i,j,n}\frac{\alpha r_{ijn}^{(\DNN)}}{\alpha r_{ijn}^{(\DNN)}+\beta c_{ijn}}\frac{1}{c_{ijn}}\tv{i}{j}{k}{n} \notag \\
    & + \sum_{i,j,n}\left(\alpha\sum_{k}\frac{\lambda_{ijk,n}^{2}}{t_{ik,n}v_{kj,n}}+\frac{\beta}{r_{ijn}^{(\DNN)}}\right)|y_{ijn}|^{2} \notag \\
    & - \sum_{i,j,n}\left[\frac{\alpha r_{ijn}^{(\DNN)}}{\alpha r_{ijn}^{(\DNN)}+\beta c_{ijn}} + \log\left(\frac{\alpha}{c_{ijn}}+\frac{\beta}{r_{ijn}^{(\DNN)}}\right)\right]. \notag \\  
\end{align}
Setting the partial derivatives of $\L^{+}_{\proposed}$ with respect to $t_{ik,n}$ and $v_{kj,n}$ equal to zero yields
\begin{align}
    t_{ik,n}
    = \left[{\frac{\displaystyle\sum_{j}\frac{\lambda_{ijk,n}^{2}}{v_{kj,n}}|y_{ijn}|^{2}}{\displaystyle\sum_{j}\frac{r_{ijn}^{(\DNN)}}{\alpha r_{ijn}^{(\DNN)}+\beta c_{ijn}}\frac{v_{kj,n}}{c_{ijn}}}}\right]^{\frac{1}{2}},
    \label{eq:proposed-method/NMF-updates/stationary_t} \\
    v_{kj,n}
    = \left[{\frac{\displaystyle\sum_{i}\frac{\lambda_{ijk,n}^{2}}{t_{ik,n}}|y_{ijn}|^{2}}{\displaystyle\sum_{i}\frac{r_{ijn}^{(\DNN)}}{\alpha r_{ijn}^{(\DNN)}+\beta c_{ijn}}\frac{t_{ik,n}}{c_{ijn}}}}\right]^{\frac{1}{2}}.
    \label{eq:proposed-method/NMF-updates/stationary_v}
\end{align}
By substituting the equality conditions \eqref{eq:proposed-method/NMF-updates/tangent-equality_hold} and \eqref{eq:proposed-method/NMF-updates/Jensen-equality_hold} into \eqref{eq:proposed-method/NMF-updates/stationary_t} and \eqref{eq:proposed-method/NMF-updates/stationary_v}, we obtain the following update rules:
\begin{align}
    t_{ik,n}
    & \leftarrow t_{ik,n}\left[{\frac{\displaystyle\sum_{j}\frac{v_{kj,n}}{\left(\tv{i}{j}{k'}{n}\right)^{2}}|y_{ijn}|^{2}}{\displaystyle\sum_{j}\frac{v_{kj}}{\left(\tv{i}{j}{k'}{n}\right)^{2}}\tilde{r}_{ijn}}}\right]^{\frac{1}{2}},
    \label{eq:proposed-method/NMF-updates/update_t} \\
    v_{kj,n}
    & \leftarrow v_{kj,n}\left[{\frac{\displaystyle\sum_{i}\frac{t_{ik,n}}{\left(\tv{i}{j}{k'}{n}\right)^{2}}|y_{ijn}|^{2}}{\displaystyle\sum_{i}\frac{t_{ik,n}}{\left(\tv{i}{j}{k'}{n}\right)^{2}}\tilde{r}_{ijn}}}\right]^{\frac{1}{2}}.
    \label{eq:proposed-method/NMF-updates/update_v}
\end{align}

\subsubsection{Update Rule of $\matf{W}_{i}$}
\label{sec:proposed-method/estimation-of-demixing-matrix}
Since the cost function \eqref{eq:proposed-method/formulation/cost} is an IP-applicable form, we can reuse the update rule \eqref{eq:conventional-methods/ILRMA/definition-of-U} merely by replacing $r_{ijn}^{(\NMF)}$ with $\tilde{r}_{ijn}$.
The PB technique is used as in ILRMA and IDLMA.

\subsubsection{Summary of Update Rules}
The entire separation process of the proposed method is shown in Algorithm~\ref{alg:proposed-method/update-algorithm}, where $L$ denotes the number of DNN-based source model updates and $L'$ denotes that of NMF-based source model and demixing matrix updates per DNN-based source model update.

\begin{figure}[!t]
\begin{algorithm}[H]
    \caption{Iterative algorithm of proposed method}
    \label{alg:proposed-method/update-algorithm}
    \begin{algorithmic}[1]
    \REQUIRE $\matf{X}_{1},\ldots,\matf{X}_{M}$, $\DNN_{1},\ldots,\DNN_{N}$
    \ENSURE $\matf{Y}_{1},\ldots,\matf{Y}_{N}$
    \FOR{$l=1,\ldots,L$}
        \FOR{all source index $n$}
            \STATE Update DNN source model $\matf{R}_{n}^{(\DNN)}$ by \eqref{eq:conventional-methods/IDLMA/estimation_IDLMA} and \eqref{eq:conventional-methods/IDLMA/flooring_IDLMA}
            \STATE Update source model $\tilde{\matf{R}}_{n}$ by \eqref{eq:proposed-method/formulation/definition-of-r}
        \ENDFOR
        \FOR{$l'=1,\ldots,L'$}
            \FOR{all source index $n$}
                \STATE Update NMF source model $\matf{R}_{n}^{(\NMF)}$ by \eqref{eq:proposed-method/NMF-updates/update_t} and \eqref{eq:proposed-method/NMF-updates/update_v}
                \STATE Update source model $\tilde{\matf{R}}_{n}$ by \eqref{eq:proposed-method/formulation/definition-of-r}
            \ENDFOR
            \FOR{all frequency bin $i$ and source index $n$}
                \STATE Update $\vecf{w}_{in}$ by \eqref{eq:conventional-methods/ILRMA/update_w1} and \eqref{eq:conventional-methods/ILRMA/update_w2} with \eqref{eq:conventional-methods/ILRMA/definition-of-U}
            \ENDFOR
            \FOR{all frequency bin $i$ and time frame $j$}
               \STATE Update $\vecf{y}_{ij}$ by \eqref{eq:conventional-methods/formulation/demixing}
            \ENDFOR
            \FOR{all frequency bin $i$ and time frame $j$}
                \STATE Apply PB technique to $\vecf{y}_{ij}$ by \eqref{eq:conventional-methods/ILRMA/projection-back} with \eqref{eq:conventional-methods/formulation/solution-of-d}
            \ENDFOR
        \ENDFOR
    \ENDFOR
    \end{algorithmic}
\end{algorithm}
\end{figure}

\section{Experimental Evaluation}
\label{sec:experiment}
\begin{figure}[tb]
    \centering
    \includegraphics[width=\linewidth]{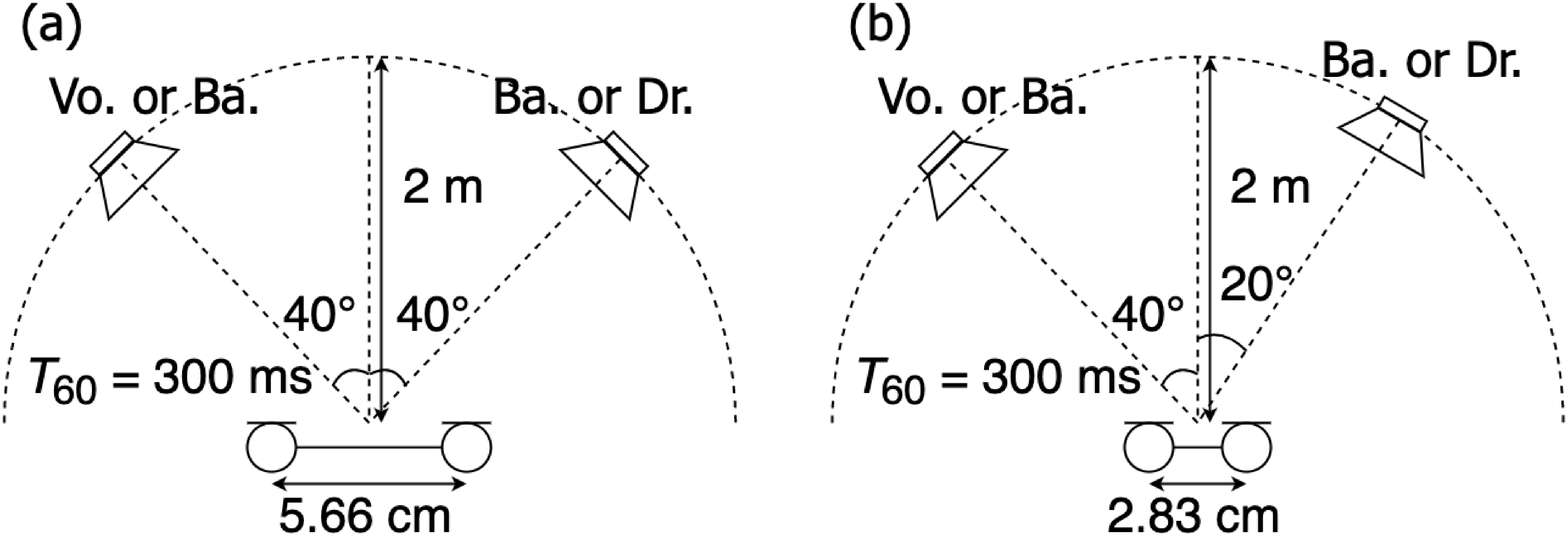}
    \caption{Recording conditions for impulse response.}
    \label{fig:experimental-evaluation/experimental-setting/recording-condition}
\end{figure}
\begin{figure*}[tb]
    \begin{minipage}[t]{0.33\hsize}
        \centering
        \includegraphics[width=\linewidth]{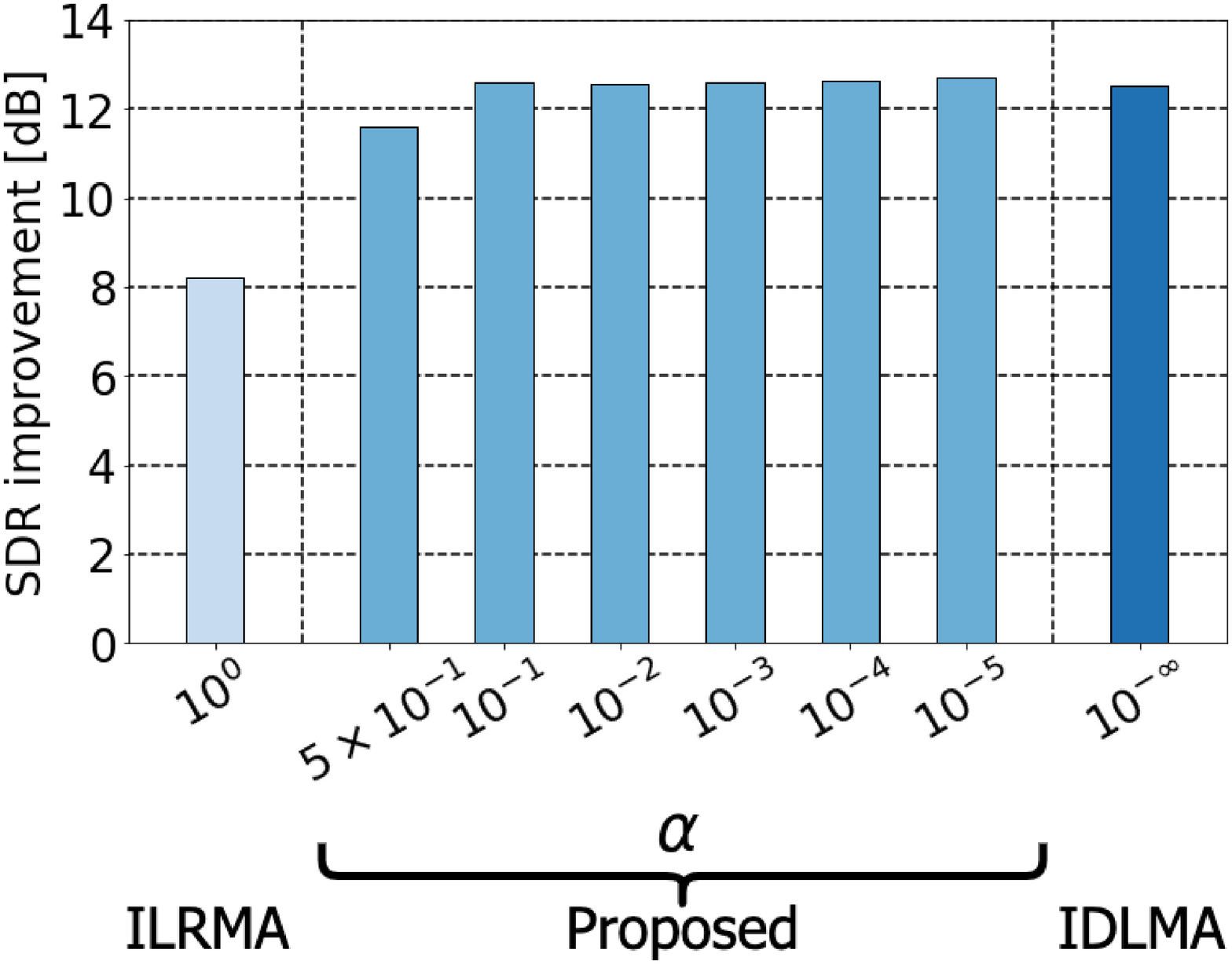}
        {(a) Vo./Ba.}
    \end{minipage}
    \begin{minipage}[t]{0.32\hsize}
        \centering
        \includegraphics[width=\linewidth]{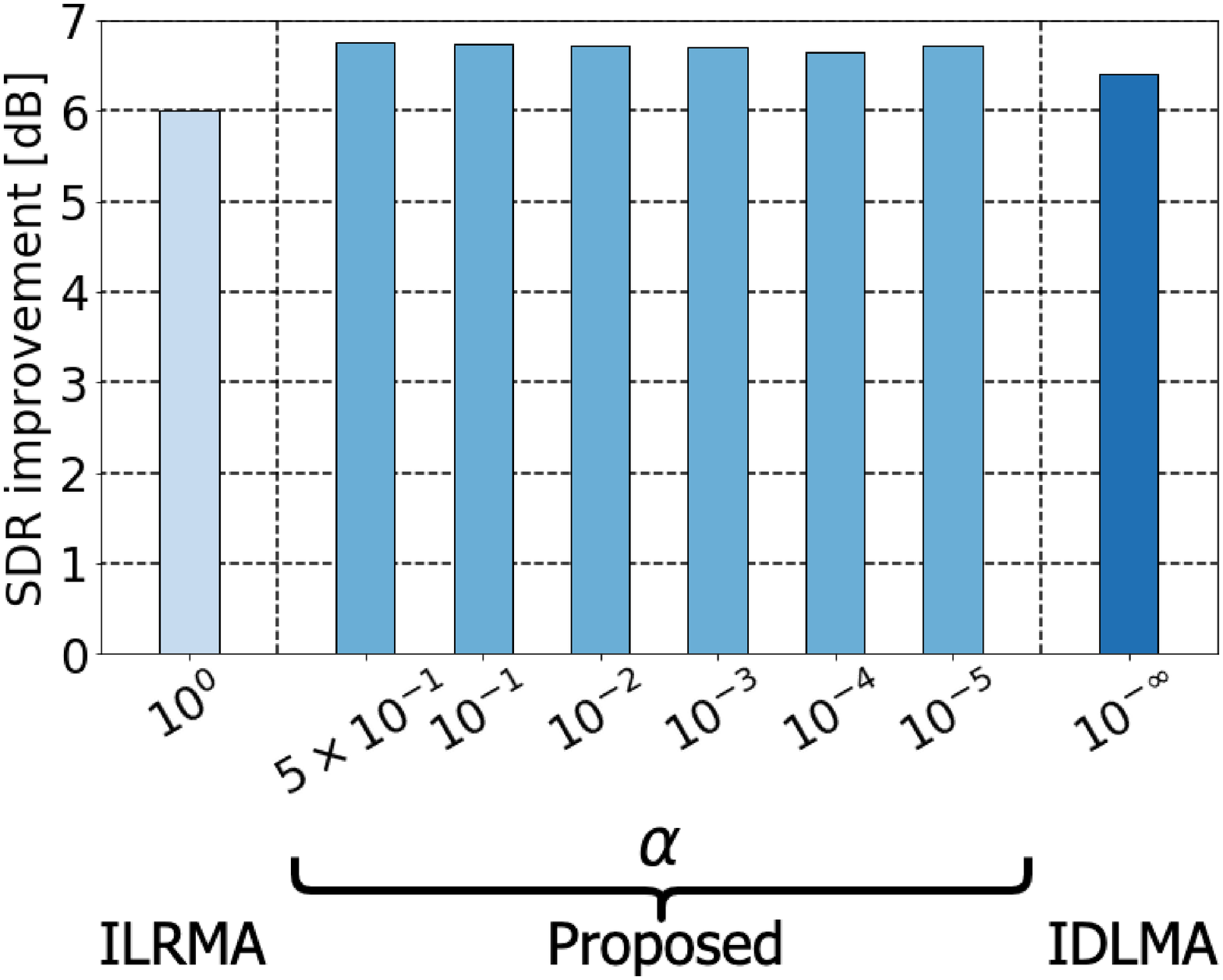}
        {(b) Ba./Dr.}
    \end{minipage}
    \begin{minipage}[t]{0.33\hsize}
        \centering
        \includegraphics[width=\linewidth]{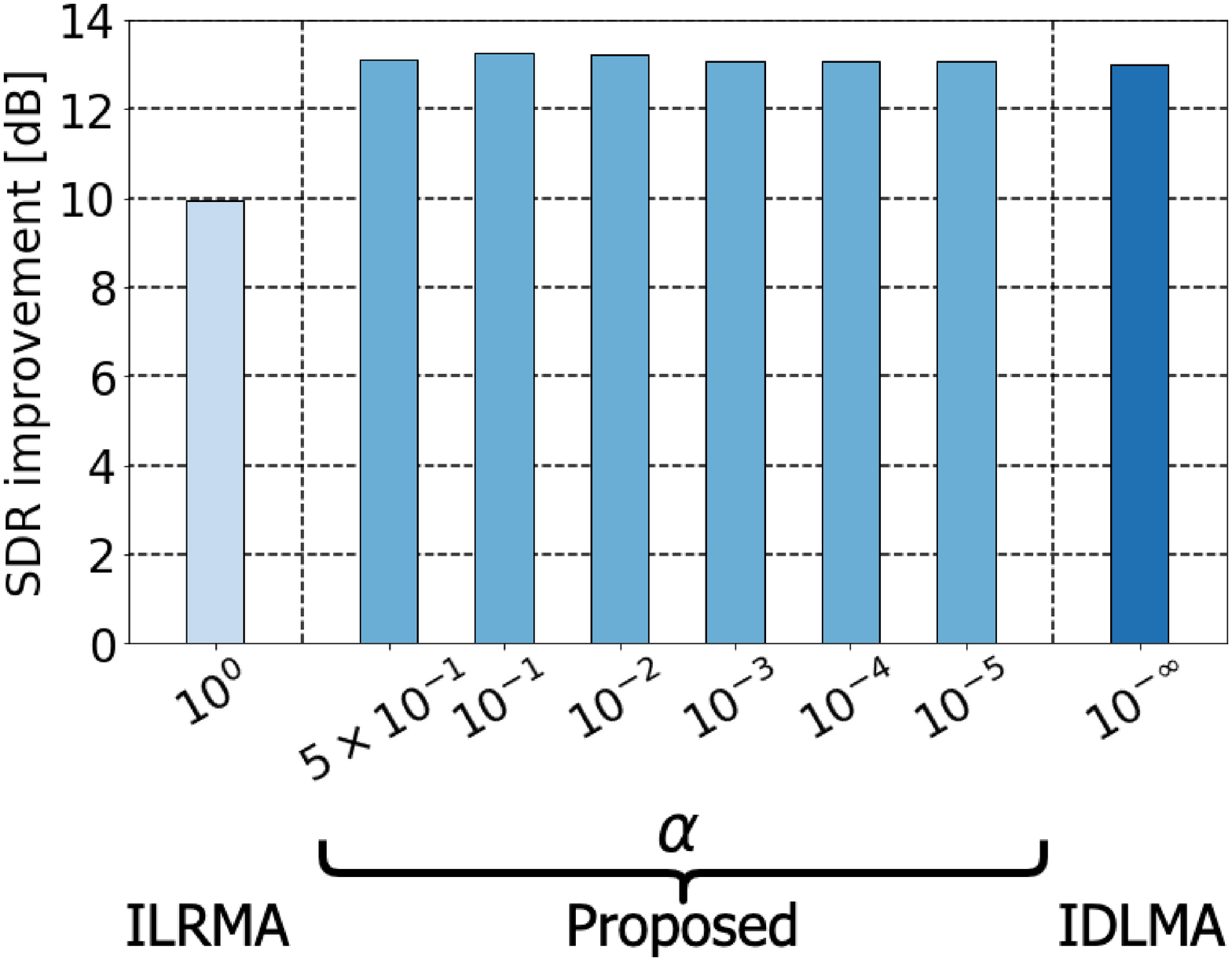}
        {(c) Vo./Dr.}
    \end{minipage}
    \caption{SDR improvements of proposed method with varying $\alpha$ and of conventional methods.}
    \label{fig:experimental-evaluation/results/SDR-improvement}
\end{figure*}
\begin{figure*}[tb]
    \centering
    \begin{minipage}[t]{0.48\hsize}
        \centering
        \includegraphics[width=\linewidth]{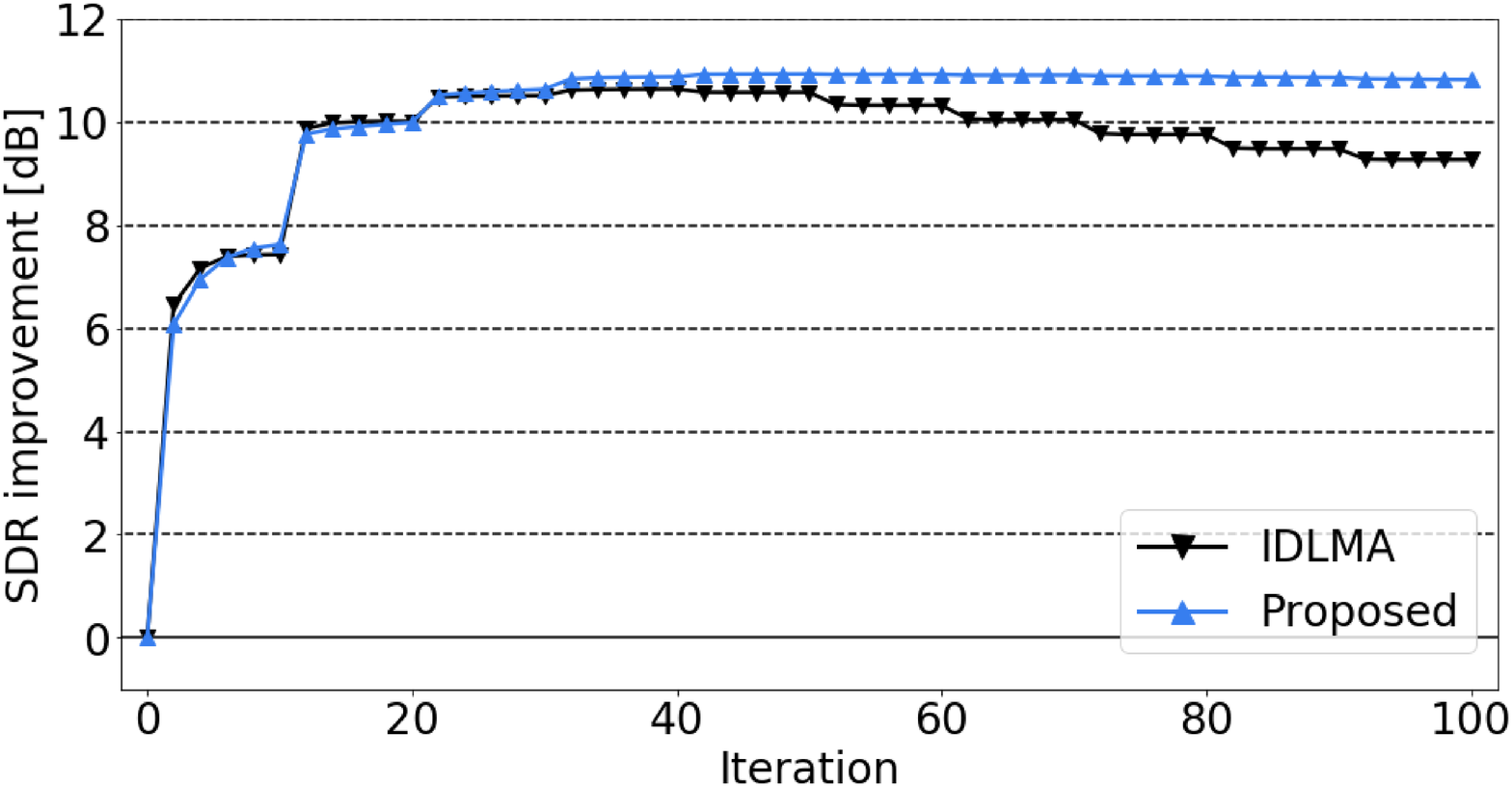} \\
        {(a) \texttt{Girls Under Glass - We Feel Alright}.}
    \end{minipage}
    \begin{minipage}[t]{0.48\hsize}
        \centering
        \includegraphics[width=\linewidth]{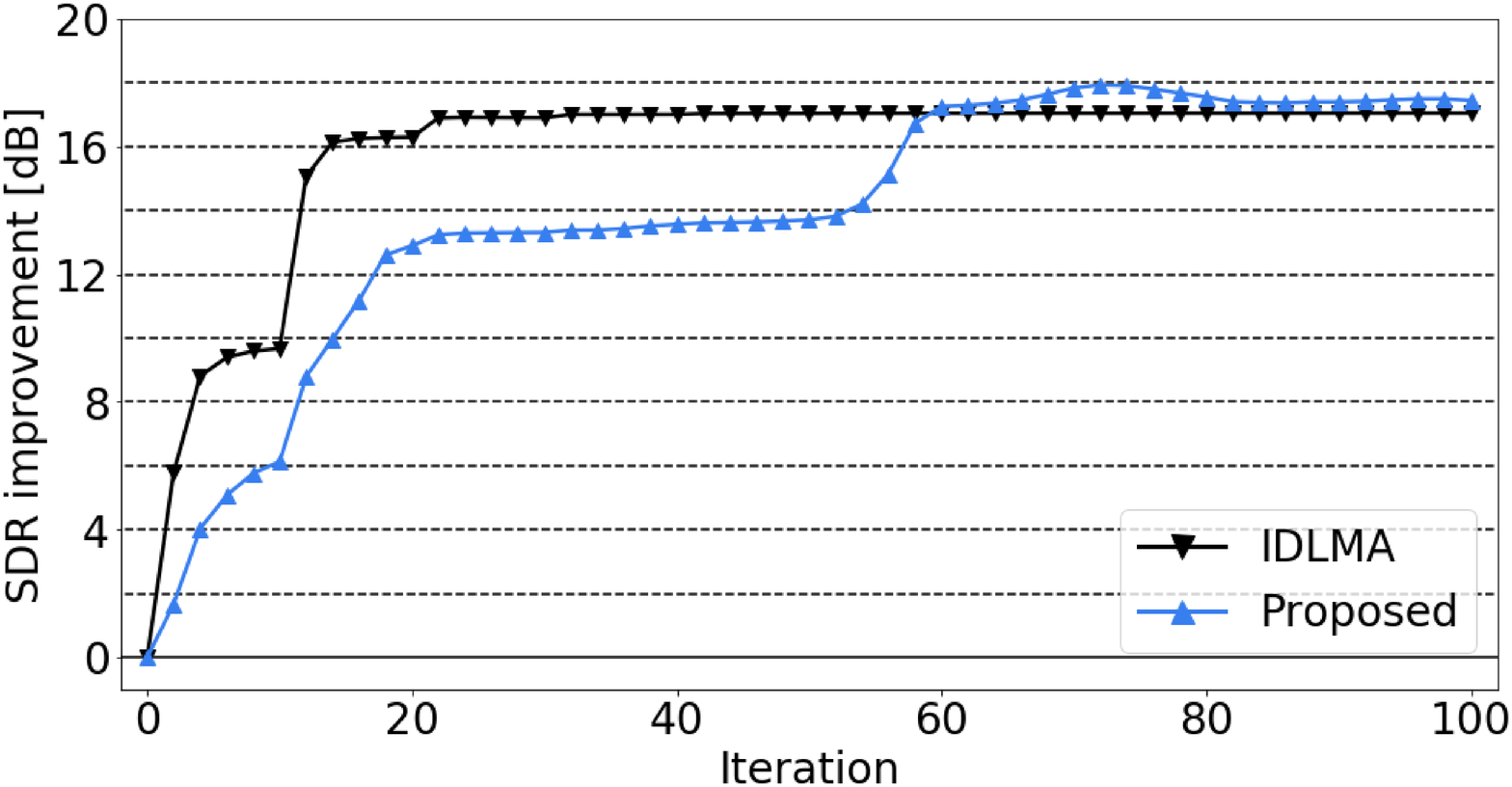} \\
        {(b) \texttt{James May - Don't Let Go}.}
    \end{minipage}
    \caption{Examples of SDR improvements of IDLMA and of proposed method for Ba./Dr.}
    \label{fig:experimental-evaluation/results/[bass,drums]_SDRi}
\end{figure*}
\subsection{Experimental Setting}
\label{sec:experimental-evaluation/experimental-setting}
To evaluate the effectiveness of the proposed method, we conducted a multichannel music source separation experiment using the DSD100 dataset~\cite{liutkus20172016}.
We downsampled the audio signals to $8$ kHz and used the Hamming window of $512$ ms with a shift length of $256$ ms for STFT.

As the test data, we created multichannel mixtures by convolving dry sources and the E2A impulse response (reverberation time is $300$ ms) in the RWCP database~\cite{nakamura2000acoustical}.
The dry sources were the $30$- to $60$-s segments of the top $25$ songs in the \texttt{test} set in alphabetical order.
We prepared three pairs of instruments: vocal and bass (Vo./Ba.), bass and drums (Ba./Dr.), and vocal and drums (Vo./Dr.).
Fig.~\ref{fig:experimental-evaluation/experimental-setting/recording-condition} shows the two recording conditions, and the number of the multichannel mixtures for each instrument pair was $50$.

We compared the proposed method with ILRMA~\cite{kitamura2016determined} and IDLMA~\cite{makishima2019independent}.
In accordance with the experimental conditions used in \cite{kitamura2016determined} and \cite{makishima2019independent},
we set $L=10$ and $L'=10$ for both IDLMA and the proposed method.
Since ILRMA does not use the DNNs, its NMF parameters and demixing matrix were updated for $100$ iterations, i.e., we set the number of NMF parameter updates to be the same as in the proposed method.
In accordance with \cite{makishima2019independent}, the number of bases for NMF was $K=20$ for both ILRMA and the proposed method.
The source-specific DNN consists of five fully connected (FC) blocks, each of which consists of an FC layer with $2048$ hidden units and a rectified linear unit (ReLU) nonlinearity.
A dropout layer with a drop rate of $0.3$ was placed after each ReLU in the FC blocks except for the last one.
As described in Section~\ref{sec:proposed-method/pretraining-and-separation-algorithm/DNN-training-and-inference}, the proposed method can use the DNNs trained in the same manner as in IDLMA, and we used the same trained DNNs for IDLMA and the proposed method.
In the proposed method, we varied $\alpha$ from $5\times10^{-1}$ to $1\times10^{-5}$ with the constraint of $\alpha+\beta=1$.

\subsection{DNN Training}
We used all $50$ songs in the \texttt{dev} set of the DSD100 dataset for DNN training, and the bottom $25$ songs in alphabetical order in the \texttt{test} set for the validation.
During the DNN training, as described in Section~\ref{sec:conventional-methods/IDLMA}, single-channel noisy mixtures were created.
The mixtures were created as described in~\cite{hasumi2021empirical}.
The DNNs were trained for $2000$ epochs on an Adadelta~\cite{zeiler2012adadelta} optimizer with a batch size of $128$.
The learning rate was set to $1.0$ with a weight of $l^{2}$ regularization of $10^{-5}$.
We clipped the norm of the gradients of the DNN parameters before the parameter updates so that their $l^{2}$ norms were less than or equal to $10$.
The other hyperparameters were set as $\delta=10^{-5}$ and $\varepsilon=10^{-1}$, following~\cite{makishima2019independent}.

\subsection{Results}
\label{sec:experimental-evaluation/results}
Fig.~\ref{fig:experimental-evaluation/results/SDR-improvement} shows the average source-to-distortion ratio (SDR) improvements over $50$ mixtures of test data for each instrument pair, which were computed with the BSSEval toolbox~\cite{vincent2006performance}.
Compared with ILRMA, IDLMA provided SDR improvements of more than $3$ dB higher in Vo./Ba. and Vo./Dr. separations.
However, for Ba./Dr., IDLMA provides an SDR improvement of only $0.4$ dB higher than ILRMA, which is lower than for the other instrument pairs.
This may be because the spectrograms of bass and drums are likely to be of low-rank, which fits the assumption of NMF.
On the other hand, the proposed method outperformed IDLMA with the appropriate choice of $\alpha$ for any instrument pair, showing the effectiveness of the proposed method.

Fig.~\ref{fig:experimental-evaluation/results/[bass,drums]_SDRi} shows the average SDR improvement at each iteration for IDLMA and the proposed method with $\alpha=5\times10^{-1}$.
The results were for the Ba./Dr. mixture and averaged over the two instruments.
In IDLMA, the SDR improvements at the DNN updates increased until the $40$th iteration, but they decreased at subsequent iterations, as shown in Fig.~\ref{fig:experimental-evaluation/results/[bass,drums]_SDRi}(a).
Although most of the DNN training data labeled as \texttt{bass} were performed by the electric bass guitar, the \texttt{bass} sound of this musical piece was performed by the synth bass.
This result shows the performance degradation caused by the timbral discrepancy between the DNN training data and the data to which IDLMA was applied.
On the other hand, the SDR improvements of the proposed method did not decrease even after approximately the $40$th iteration.
This phenomenon can also be observed in Fig.~\ref{fig:experimental-evaluation/results/[bass,drums]_SDRi}(b).
The \texttt{drums} sound of this musical piece was performed by the conga, which was not included in the training data.
Although the average SDR improvements of IDLMA seemed to reach the upper performance limit, the proposed method gave higher average SDR improvements in the later iterations.
These results show that the introduction of the source model independent of the DNN training data reduces the performance degradation caused by the timbral gap.

\section{Conclusion}
\label{sec:conclusion}
We proposed the PoSM by combining the DNN- and NMF-based source models, which are respectively used in IDLMA and ILRMA.
The DNN-based part represents the components similar to the training data, and the NMF-based part represents the components independent of the training data.
We introduced the PoSM into IDLMA to develop the PoSM-based IDLMA and derived the computationally efficient separation algorithm of its parameters.
For the NMF parameters, we derived the convergence-guaranteed iterative algorithm based on the MM algorithm.
Through the multichannel music source separation experiments, we showed the effectiveness of the proposed PoSM-based IDLMA.
Furthermore, we showed that the use of the PoSM reduces the performance degradation caused by the timbral gap between the DNN training data and the sounds to which the proposed method is applied.

\section*{Acknowledgment}
This work was supported by JSPS-CAS Joint Research Program, Grant number JPJSBP120197203, and JSPS KAKENHI Grant numbers 19K20306, 19H01116, and 17H06101.

\end{document}